%% file: main.tex
\documentclass[sigconf]{acmart}
\AtBeginDocument{%
  \providecommand\BibTeX{{%
    \normalfont B\kern-0.5em{\scshape i\kern-0.25em b}\kern-0.8em\TeX}}}

\copyrightyear{2024} 
\acmYear{2024}
 \setcopyright{rightsretained}
  \acmConference[IDC '24]{Interaction Design and Children}{June 17--20, 2024}{Delft, Netherlands}
   \acmBooktitle{Interaction Design and Children (IDC '24), June 17--20, 2024, Delft, Netherlands}
   \acmDOI{10.1145/3628516.3655809}
   \acmISBN{979-8-4007-0442-0/24/06}

\usepackage[utf8]{inputenc}
\usepackage{cleveref}
\usepackage{listings}

\newcommand{\system}{ContextQ}

\begin{document}

\title[\system{}]{\system{}: Generated Questions to Support Meaningful Parent-Child Dialogue While Co-Reading}

\author{Griffin Dietz Smith}
\affiliation{
  \institution{Apple}
  \city{Seattle}
  \state{WA}
  \country{USA}
}
\email{griffind@apple.com}

\author{Siddhartha Prasad}
\affiliation{
  \institution{Brown University}
  \city{Providence}
  \state{RI}
  \country{USA}
}
\email{siddhartha_prasad@brown.edu}

\author{Matt J. Davidson}
\affiliation{
  \institution{Apple}
  \city{Austin}
  \state{TX}
  \country{USA}
}
\email{matt_davidson@apple.com}

\author{Leah Findlater}
\affiliation{
  \institution{Apple}
  \city{Seattle}
  \state{WA}
  \country{USA}
}
\email{lfindlater@apple.com}

\author{R. Benjamin Shapiro}
\affiliation{
  \institution{Apple}
  \city{Seattle}
  \state{WA}
  \country{USA}
}
\email{nerd@apple.com}

\renewcommand{\shortauthors}{Dietz Smith, et al.}

\begin{abstract}
\input{00-abstract}
\end{abstract}

%



\maketitle

\input{01-intro}
\input{02-rw}

\input{03-system-design}
\input{04-method}
\input{05-results}
\input{06-discussion}
\input{07-conclusion}
\input{08-selection-of-children}
\begin{acks}
We would like to thank Katie Van Sluys, Katie Morrow, Gabriella Meyers, and Megan Kinsey for contributing their expertise to rubric development and evaluation.
\end{acks}

\bibliographystyle{ACM-Reference-Format}
\bibliography{references}

\input{09-appendix}

\end{document}

%% file: 00-abstract.tex
Much of early literacy education happens at home with caretakers reading books to young children. Prior research demonstrates how having dialogue with children during co-reading can develop critical reading readiness skills, but most adult readers are unsure if and how to lead effective conversations. We present \system{}, a tablet-based reading application to unobtrusively present auto-generated dialogic questions to caretakers to support this dialogic reading practice. An ablation study demonstrates how our method of encoding educator expertise into the question generation pipeline can produce high-quality output; and through a user study with 12 parent-child dyads (child age: 4--6), we demonstrate that this system can serve as a guide for parents in leading contextually meaningful dialogue, leading to significantly more conversational turns from both the parent and the child and deeper conversations with connections to the child's everyday life.

%% file: 01-intro.tex
\section{Introduction}
\label{s:intro}
Early childhood reading level is one of the most important predictors of lifetime success \cite{hernandez2011double}. Yet, over a third of children entering kindergarten lack reading readiness skills (e.g., phonemic awareness and vocabulary) and start formal schooling already below grade level \cite{boyer1991ready}. These children who start behind rarely catch up \cite{francis1996developmental, juel1988learning, torgesen1998consistency}. Much of the necessary early literacy education happens at home, through caretakers reading books with young children, but most adult readers do not realize that children learn so much more when parents and children \textit{talk} about the stories they read \cite{krcmar2014parent}. Specifically, through a style of co-reading called ``dialogic reading'' \cite{whitehurst1988accelerating, zevenbergen2003dialogic}, adults lead contextually meaningful dialogue with children by asking specific types of story-related questions that develop vocabulary, phonemic awareness, recall, expressive fluency, and the building of connections between stories and a child's life. This practice can move children ahead of peers in tests of language ability \cite{whitehurst1988accelerating}, build children’s expressive vocabulary \cite{troseth2020enhanced}, improve comprehension \cite{doyle2006promoting}, and increase their scores on measures of linguistic complexity \cite{zevenbergen2003dialogic}.

Enacting dialogic reading requires awareness of the importance of book-based conversation, knowledge of what questions are good \cite{zevenbergen2003dialogic}, and cognitive attention to generate such questions while simultaneously reading aloud, which is a difficult task for adult readers to perform. Prior work has shown that eBooks that present adults with appropriate questions for their child increase on-topic dialogue during co-reading \cite{stuckelman2021value}. However, existing systems are hard-coded with educator-written prompts, making them inflexible and unscalable \cite{troseth2020enhanced}, or they target comprehension rather than language development, and so are focused on asking specific kinds of content-based questions that are less conducive to open-ended dialogue \cite{zhang2022storybuddy}. Generating pedagogically-sound dialogic reading prompts at scale remains an open challenge.

Large language models (LLMs) present a promising new avenue for such dialogic question generation because they can use context to create questions that go beyond the immediate page (e.g., asking readers to make predictions or relate narrative events to their own experiences). However, while LLMs like GPT3 have helped make generative AI more accessible to the general public, operationalizing these models for specific applications can be challenging. Developers of LLM applications typically rely on adaptations like fine-tuning in order to overcome output quality concerns for specific tasks \cite{bommasani2021opportunities, korngiebel2021considering}, but this process is computationally expensive, requires high-quality task-specific data, and demands machine learning (ML) domain expertise \cite{bommasani2021opportunities}. For a task like dialogic question creation, determining quality in outputs also requires educational domain knowledge, so curation and annotation of new training data can necessitate dozens to hundreds of hours of expert time \cite{xu2022fantastic}. In this paper we demonstrate a new technique that is effective and orders of magnitude cheaper to instantiate.

As an alternative to fine-tuning, we draw inspiration from instructional rubrics and computational formal methods to create a self-assessing and self-correcting system. Instructional rubrics are a classroom tool in which teachers provide scoring criteria for students to self-assess their own work \cite{vasileiadou2021examining, andrade2000using, sundeen2014instructional}. By creating a rubric to identify a good dialogic question, we encode educator expertise to evaluate data quality with comparatively little expert time and without curating a dataset. Then, to mimic self-evaluative practices for applying these rubrics within our system, we turn to counter-example guided inductive synthesis (CEGIS), a formal methods technique \cite{cegis, CevskaProbCEGIS}. Traditionally, this method uses test cases to self-supervise and self-correct outputs in program synthesis. By building a similar layer into a question generation system that uses the rubric scores to give targeted feedback for improvement, we can identify and generate high quality questions without fine-tuning. 

In this work, we describe the design, development, and user evaluation of \system{}, an automated LLM-enabled system for supporting dialogic questioning during parent-child co-reading. \system{} has two components, a question generation module and a tablet-based reading interface. The question generation module uses an LLM to generate dialogic questions and applies a rubric to identify suitable outputs, providing feedback to the LLM as needed. The tablet-based reading interface then presents the selected questions unobtrusively to parents during co-reading to spur contextually meaningful dialogue. A qualitative evaluation of \system{} investigates its impact on reading-time conversation and how parents use it as a tool in practice.

Specifically, our research contributes:
\begin{itemize}
    \item an approach for improving LLM output quality by applying expert-informed rubrics in a self-supervising system
    \item \system{}, an LLM-supported tablet application for dialogic questioning that leverages the rubric-based approach for question generation
    \item a qualitative evaluation with 12 parent-child dyads that demonstrates how \system{} can be a supportive tool to guide parents in leading deeper conversations that fit learning objectives for their child (aged 4--6)
\end{itemize}

%% file: 02-rw.tex
\section{Related Work}
To situate these contributions, we review literature on dialogic reading as an educational practice, systems to support dialogic reading, and methods for question generation.

\subsection{Dialogic Reading as an Educational Practice}
\label{s:rw-dq}
Children's early experiences with books play an important role in reading readiness. Picture book reading can help children develop skills related to vocabulary, phonemic awareness, print meaning, narrative structure, and more \cite{duursma2008reading}. But \textit{how} we read to children matters just as much as how frequently we read to them \cite{whitehurst1988accelerating, beck2001text}.

Dialogic reading is a style of co-reading in which the adult reader encourages the child to talk about the picture book while reading and in doing so models progressively more sophisticated language \cite{whitehurst1988accelerating}. Children who have been read to following this dialogic technique can move significantly ahead of their peers on tests of expressive language ability even after just a few weeks \cite{whitehurst1988accelerating}. They will also talk more often while reading and for longer than children who have been read to in a traditional manner \cite{whitehurst1988accelerating}. Critically, these findings have held time and again across children from different countries, learning in different settings (e.g., home or daycare), and coming from a wide range of economic backgrounds \cite{whitehurst1988accelerating, valdez1992accelerating, arnold1994accelerating, lim2002facilitating}.

The most popular strategy for facilitating dialogic reading follows the CROWD acronym \cite{whitehurst1994outcomes, zevenbergen2003dialogic}, which outlines the types of prompts a parent might use to foster meaningful conversation. These question types are completion, recall, open-ended, wh-, and distancing (see Table \ref{t:crowd}). Each of these types of prompts has a specific educational objective, and successful dialogic reading practice will include questions of different types to advance each of these pedagogical goals. Notably, other recent strategies to facilitate parent-child conversation focus more heavily on abstract, open-ended discussion, but these methods were not developed for reading contexts \cite{leech2018brief, leech2021intervention}.

\begin{table*}[] 
\renewcommand{\arraystretch}{1.5}
\begin{tabular}{lp{0.27\textwidth}p{0.27\textwidth}p{0.27\textwidth}} 
\textbf{Question Type} & \textbf{Description} & \textbf{Educational Objective} & \textbf{Example Question} \\ \hline
\textbf{C}ompletion & Ask children to complete a sentence or phrase from the book being read. & Build phonemic awareness and introduce the structure of language. & I’ll huff, and I’ll puff, and I’ll blow the house \_\_\_\_. \\
\textbf{R}ecall & Ask children about story plot, typically spanning multiple pages. & Build story comprehension, typically around sequences of events or overall themes. & Which house couldn’t the Big Bad Wolf blow down? \\
\textbf{O}pen ended & Encourage children to express their own ideas and opinions about the story. & Discuss the story beyond the content of the text (i.e., to make predictions or inferences). & How do you think the pigs felt when the wolf tried to get them? \\
\textbf{W}h- & Solicit descriptive details, and begin with \textit{what}, \textit{when}, \textit{who}, \textit{where} or \textit{why}. & Develop vocabulary or evaluate comprehension of a page. & What did the first pig make his house out of? \\
\textbf{D}istancing & Relate the story to children's lives and lived experiences. & Build connections between the story and the child's own life. & What's a time that someone broke something of yours? How did you feel?
\end{tabular}  
\caption{CROWD identifies five types of questions that could initiate dialogue. Each question type has a specific educational objective \cite{whitehurst1988accelerating,zevenbergen2003dialogic}. The last column includes examples of each prompt type for the story, ``The Three Little Pigs.''}
\label{t:crowd}
\end{table*}

While CROWD describes the types of prompts that could foster meaningful co-reading conversation, additional characteristics of the question itself can influence its suitability in a given context. In pedagogical practice, an authentic question is loosely defined as one without a prescribed answer, motivated by a genuine desire to learn something from the respondent \cite{Nystrand_1997, Schaffalitzky2022WhatMA}. More authentic questions also typically map to higher levels in taxonomies that categorize questions by the complexity of thought and expressive language needed to respond \cite{flynn2011, Costa_2001, costa1989techniques}. By nature of their uncertainty and complexity, these questions are more germane to dialogue, and the number of authentic questions asked by educators has functioned as a heuristic by which to quantify the quality of classroom discussion \cite{Nystrand_1997, nystrand2003questions}. 

It is not enough just to ask a question. Asking the right question means asking a question that fosters contextually meaningful conversation, aligns with pedagogical goals, and has the correct degree of complexity and authenticity. 

\subsection{Systems to Support Dialogic Reading} \label{s:rw-systems}
Given the nuances of asking ``good'' questions, leading dialogic questioning can be a rather demanding task for the adult reader. To do so, they must know a) that asking questions is a valuable practice and b) what makes a good question. To address these challenges, researchers developed a variety of strategies aimed at supporting family dialogic reading practices.

Early intervention studies brought parents into lab settings for in-person training led by highly-skilled teachers \cite{whitehurst1994outcomes, mol2008added}. In practice, these trainings pose access challenges to parents with limited time and resources \cite{arnold1994accelerating, hindman2016closing}. Researchers have developed video-taped trainings for caretakers to watch at home, but doing so still requires awareness and time on the part of the parent \cite{arnold1994accelerating, strouse2013effective}.

eBooks present a promising opportunity to encourage parent-child engagement through appropriate interaction design \cite{troseth2020enhanced, korat2010new}. Prior research has explored the use of an on-screen agent to drive dialogic questioning \cite{strouse2013effective, troseth2020enhanced, xu2023rosita}. Strouse et al. developed an eBook for 3-year-old children with pre-recorded videos of a preschool teacher that would periodically appear in the top corner of the screen to ask a question \cite{strouse2013effective}, and Troseth et al. developed an enhanced eBook fro children aged 3--5 in which a character from the book appeared on-screen to ask the child questions \cite{troseth2020enhanced}. More recently, Xu et al. created a conversational agent to ask questions during reading with bilingual families with children aged 3--6 \cite{xu2023rosita}. These studies revealed that agent-driven dialogue in eBooks can lead to increased conversational turns and lexical diversity (i.e., number of unique words) between a parent and child \cite{troseth2020enhanced}, increases in parent's dialogic question-asking on pages without character-driven prompts \cite{troseth2020enhanced}, and improvements in children's story comprehension and story vocabulary \cite{strouse2013effective}.

Rather than relying on an agent to ask questions, StoryVisit instead presented questions to a remote adult co-reader \cite{raffle2011hello}. This interface supported children under age 6 and long-distance adults in reading children’s books together through a videochat co-reading interface, with the intention of increasing the duration of video calls \cite{raffle2011hello}. However, the authors found the presented questions in this app were only used about 5\% of the time, which they argued was due to the presence of too many other features \cite{raffle2011hello}.

While the above interfaces all utilized hard-coded questions for a handful of stories, StoryBuddy is an interface for presenting ML-generated questions to children aged 3--8 while reading \cite{zhang2022storybuddy}. It incorporates an AI agent that can read to the child, ask questions generated from a question answering model, and assess the correctness of responses. StoryBuddy also has a corresponding parent co-reading mode which allows the parent to take over any or all of these activities. However, the emphasis of StoryBuddy is in developing reading comprehension through the asking and evaluation of questions with a clear correct answer and on building a system that can minimize parent involvement as needed \cite{zhang2022storybuddy}, as opposed to focusing on developing narrative sense through rich parent-child dialogue, often spurred by open-ended and distancing questions that do not necessarily have a distinctly correct response. 

This prior research demonstrates the promise of utilizing eBooks to drive parent-child dialogue \cite{troseth2020enhanced, raffle2011hello} and the potential for machine learning to scale this approach through question generation \cite{zhang2022storybuddy}. However, there are no systems to date aimed at automatically generating and presenting dialogic questions to adult readers in co-reading situations.

\subsection{Methods for Question Generation}
\label{s:rw-llms-and-rubrics}
Question answer generation (QAG) is a field in natural language processing aimed at automatically generating both questions and answers from a body of text. Work in the QAG space primarily focuses on improving accuracy of generated questions and answers using rule-based (e.g.,  \cite{curto2011exploring, labutov2015deep}) or neural-network-based (e.g., \cite{du2017learning, dong2019unified, tang2017question}) models. This accuracy measures similarity to human-generated questions, but does not generally emphasize pedagogical value. 

The model underlying StoryBuddy, on the other hand, focuses on reading comprehension and was trained on a FairytaleQA, a dataset of over 10,000 questions and answers hand-written by educational experts \cite{zhang2022storybuddy, yao2021ai, xu2022fantastic}. While more recent QAG datasets have begun exploring open-ended and ``unanswerable'' questions \cite{choi2018quac}, we could not find evidence of work that seeks to generate authentic questions for the purpose of driving dialogue between co-readers.

Given their broad applicability, large language models (LLMs) present a promising approach to such dialogic question generation. However, these foundation models have known output quality issues that developers must address \cite{bommasani2021opportunities, korngiebel2021considering}, and there is a further set of knowledge, expertise, and experience from educators in the context of dialogic question generation that we must consider. For the most part, LLMs (like neural-network-based QAG models) are adapted for specific use-cases through fine-tuning, but fine-tuning is (often prohibitively) computationally expensive and requires a high-quality dataset \cite{bommasani2021opportunities}. While one option for question generation using LLMs would be to hire educators with the background and skillset to generate pedagogically-sound ground-truth data and then to fine-tune using that dataset, this approach would necessitate dozens to hundreds of hours of expert time, require significant computational resources, and be inflexible to changes as base models improve.

To address the limitations around data annotation, we have developed an approach to encode educator expertise into the system directly and with orders of magnitude less expert time by using a rubric. Instructional rubrics are short documents to describe the quality of a written assignment to students \cite{andrade2000using}. They encode an educator's expertise and expectations to support students' self-evaluative practices when learning to write \cite{andrade2000using, goodrich1996student, sundeen2014instructional, vasileiadou2021examining}.

For an LLM to leverage a rubric without computationally intensive fine-tuning, it must be able to self-supervise its own output. Prior work in machine learning has explored self-supervision for model training (e.g., GANs \cite{goodfellow2020generative}). Within the space of computational formal methods and program synthesis (e.g., CEGIS \cite{cegis, CevskaProbCEGIS}), self-supervision results in additional specifications to input into the synthesizer. In the context of LLMs, self-supervision often takes the form of safeguards and guardrails that accompany deployed models to detect and block potentially harmful inputs or outputs \cite{markov2023holistic, rebedea2023nemo}. Prior work has shown how fine-tuned LLMs can successfully apply rubrics to open-ended data \cite{kim2023prometheus} and how off-the-shelf models might be applied to automate deductive coding of such data \cite{Xiao_2023, sherin2013computational}, lending promise to the approach of leveraging rubrics for self-supervision. In this work, we explore how LLMs can evaluate and self-correct their own output for a task-specific application---dialogic question generation. 

%% file: 03-system-design.tex
\newcommand{\synth}{question synthesizer}
\newcommand{\recog}{suitability recognizer}
\newcommand{\Synth}{Question synthesizer}
\newcommand{\Recog}{Suitability recognizer}

\section{\system{}}

\system{} comprises two components: a question generation module and a co-reading interface.

\subsection{Question Generation Module}
\label{s:qgen}
To generate high-quality dialogic questions, we borrow the notion of instructional rubrics from writing education. Inspired by this practice, we created a rubric that encodes educator expertise for identifying a good dialogic question. To implement the self-evaluative process of applying rubrics, we took inspiration from computational formal methods to build a layer into our question generation architecture that provides self-correction via counter-examples. We demonstrate through ablation (i.e., comparison via removal of specific parts of the system architecture) how this approach yields better questions than directly generated outputs.

\subsubsection{A Rubric for Dialogic Questions}
In the broadest sense, a question is ``suitable'' to present to a family while co-reading if it could result in pedagogically meaningful dialogue between co-readers. While this suitability may be influenced by extrinsic factors (e.g., a child's age), there are aspects of prompts that intrinsically make them more likely to elicit conversation. We looked to identify these implicit characteristics and codify them through a rubric that could classify and explain suitability decisions.

Following Boyatzis' procedures for creating codebooks from theory and data \cite{gunby2011, boyatzis98}, we began this process by constructing codes reflecting relevant research (see \Cref{s:rw-dq}). In determining potential codes, three researchers had a series of discussions about the frameworks that guided dialogic reading and its desired outcomes. The initial rubric was separated by each CROWD prompt type and considered question wording, authenticity, and complexity. We then reviewed and revised these codes in the context of a supervised co-reading session between one researcher and their child. Finally, two of these researchers independently applied the candidate rubric to a set of LLM-synthesized questions. These researchers had an agreement rate (Cohen's $\kappa$) of $0.79$ and met to discuss and resolve all disagreements. These resolutions were used to produce a candidate theory-driven rubric.

We then iterated upon this theory-based rubric through four talk-aloud interviews with a literacy expert\footnote{The literacy expert is an experienced researcher and former professor of literacy focused on critical literacy and classroom curriculum, multilingual classroom communities, teacher inquiry, and collaborative teacher education.} totaling less than 3 hours. In each session, we presented this expert with text from a children's story and dialogic questions deemed suitable by the rubric. We asked them to rate each question’s likelihood of provoking meaningful contextual dialogue and to explain these decisions. We collected video and audio recordings of these interview sessions, enabling researchers to analyze sessions in detail.

We then applied iterative coding to these session recordings, looking for concepts and dimensions not captured by the extant rubric. This analysis informed new rubric iterations, which were used in subsequent sessions with the expert. Dimensions that surfaced and informed rubric revisions included the lexical structure of questions, their thematic relevance to the story, and their relevance to children's lived experiences. We achieved saturation after four sessions.

Two researchers then engaged in iterative coding, applying the rubric on a small set of questions generated from a range of children's stories and meeting to discuss and resolve identified ambiguity. These discussions led to rubric clarifications to ensure alignment with both theory and data described above. 

Finally, these researchers independently coded 116 questions synthesized from four children's stories and had an inter-rater reliability (Cohen's Kappa) of $0.87$. The final rubric considers level, authenticity, grammatical structure, relevance, and other question type-specific criteria to determine suitability (see Appendix \ref{a:rubrics} for complete rubric).

\subsubsection{Question Generation Architecture}
\begin{figure}
    \centering
    \includegraphics[width=\columnwidth]{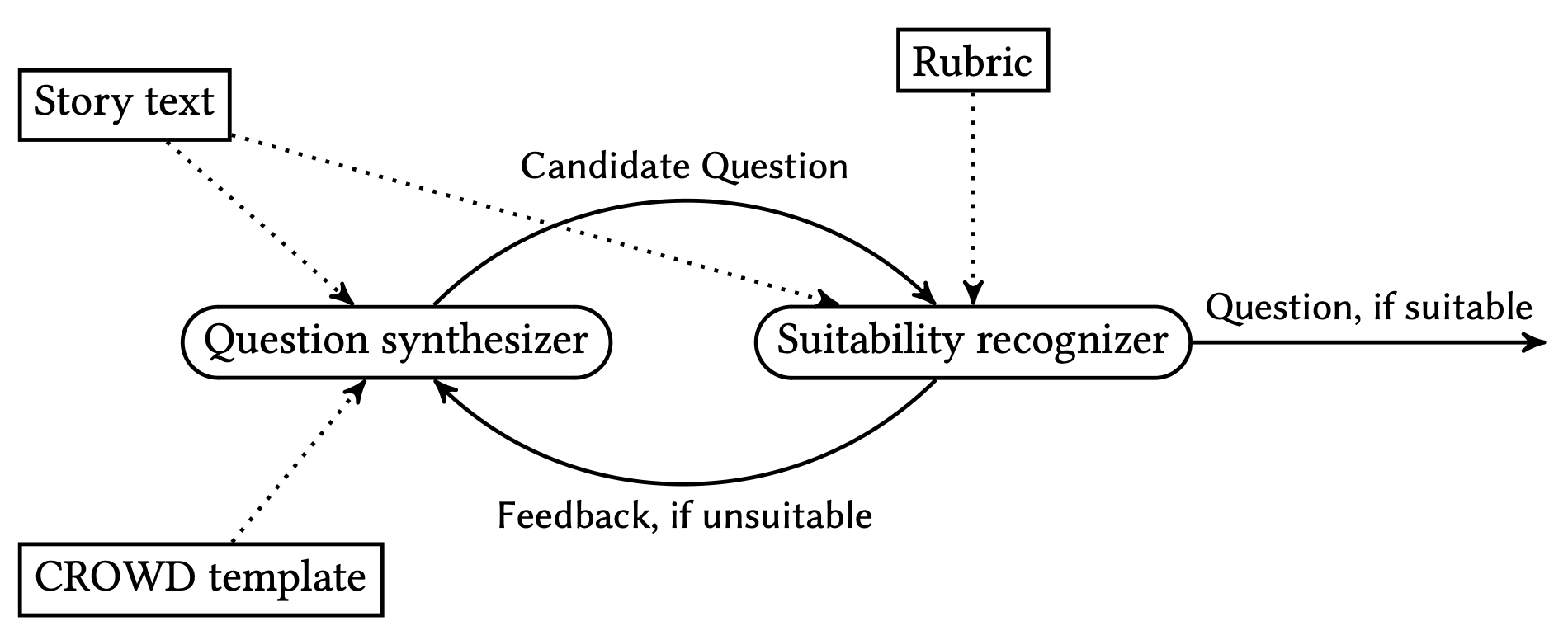}
        
        
                
        \caption{The question generation module utilizes a question synthesizer and suitability recognizer to produce high-quality dialogic questioning prompts. Dotted lines denote fixed inputs.\\}
    \label{f:prompt-loop}
\end{figure}

While generative models can be relied upon to quickly synthesize a huge number of candidate questions, this approach granted us few guarantees about the quality or correctness of these outputs. \system{} addresses this problem by applying the rubric in a two-part architecture, whose interactions are described in \Cref{f:prompt-loop}.

The \textbf{\synth{}} uses the source book text and the CROWD template to engineer LLM prompts to generate candidate dialogic questions. If provided feedback on unsuitable questions from the \recog{}, the synthesizer augments the prompt in-context \cite{liu2021pretrain}.
    
The \textbf{\recog{}} takes on the self-assessment and self-correction jobs of rubric application. It first operationalizes the dialogic questioning rubric through a series of prompts; these are fed into an LLM alongside candidate questions in order to assess output quality. When a question is recognized as unsuitable, the recognizer prompts the LLM to identify \emph{why} this decision was made. The candidate question and natural-language explanation of its unsuitability are then provided as a counter-example \cite{cegis, CevskaProbCEGIS} back to the \synth{}. 

While traditional fine-tuning methods would require dozens of hours of expert time for data annotation, we encode domain expertise in the rubric with just three hours of expert time, and feedback is incorporated without computationally expensive model training. All feedback produced by the \recog{} is acted upon by the \synth{} by means of prompt adjustment, rather than any model updates. These prompts are not only specific to the task, context, and domain, but also robust to model updates and behavior changes (i.e., as base LLMs improve, we do not need to fine-tune updated models). Additionally, because the pool of candidate questions can be cheaply increased, this synthesis-recognition loop can run for several iterations until the desired number of suitable outputs is met. We used OpenAI GPT-3.5-Turbo as the LLM in our implementation because of its combined accuracy and speed. We include example prompts in Appendix \ref{a:prompts}.

\subsubsection{Question Quality Evaluation}
\label{s:q-gen-eval}
To understand the impact of the \recog{}'s rubric-based self-assessment approach and to evaluate the quality of questions produced by \system{}'s question generation module, we compared it to an ablated architecture consisting of \textit{only} the \synth{}. In total, we generated 330 questions across five stories with half produced by each system. We presented these questions to four primary-school educators who are employees at our company. They were asked to rate the likelihood of each question to foster contextually meaningful dialogue between a parent and a child on a scale from 1 (very unlikely) to 5 (very likely). Each rater scored an equal number of questions from each system while blinded to which system produced each question. Although 3 is typically considered a neutral choice on a 5-point scale, raters articulated that they used it to mean that the question \textit{could} lead to contextually meaningful dialogue for certain families (e.g., depending on child age). In other words, the question was suitable, but depending on the family may be more or less fruitful. 

Reasoning that a question's score depends on both the generation system and the biases of the rater themself, we ran an ordinal logistic regression with assigned ratings as a dependent variable and system and rater as the independent variables. Controlling for rater, we found that questions generated by \system{} were $1.64$ times more likely to have a higher rating than those generated by the ablated architecture lacking the \recog{} (95\% CI: [$1.11$, $2.42$]). 


Of the 165 questions produced by \system{}'s question generation module, educators gave 131 of them a score of 3 or higher, representing an overall suitable question rate of $79\%$. By contrast, the ablated architecture had a suitability rate of $69\%$. 

\begin{figure*}
    \centering
    \includegraphics[width=0.8\textwidth]{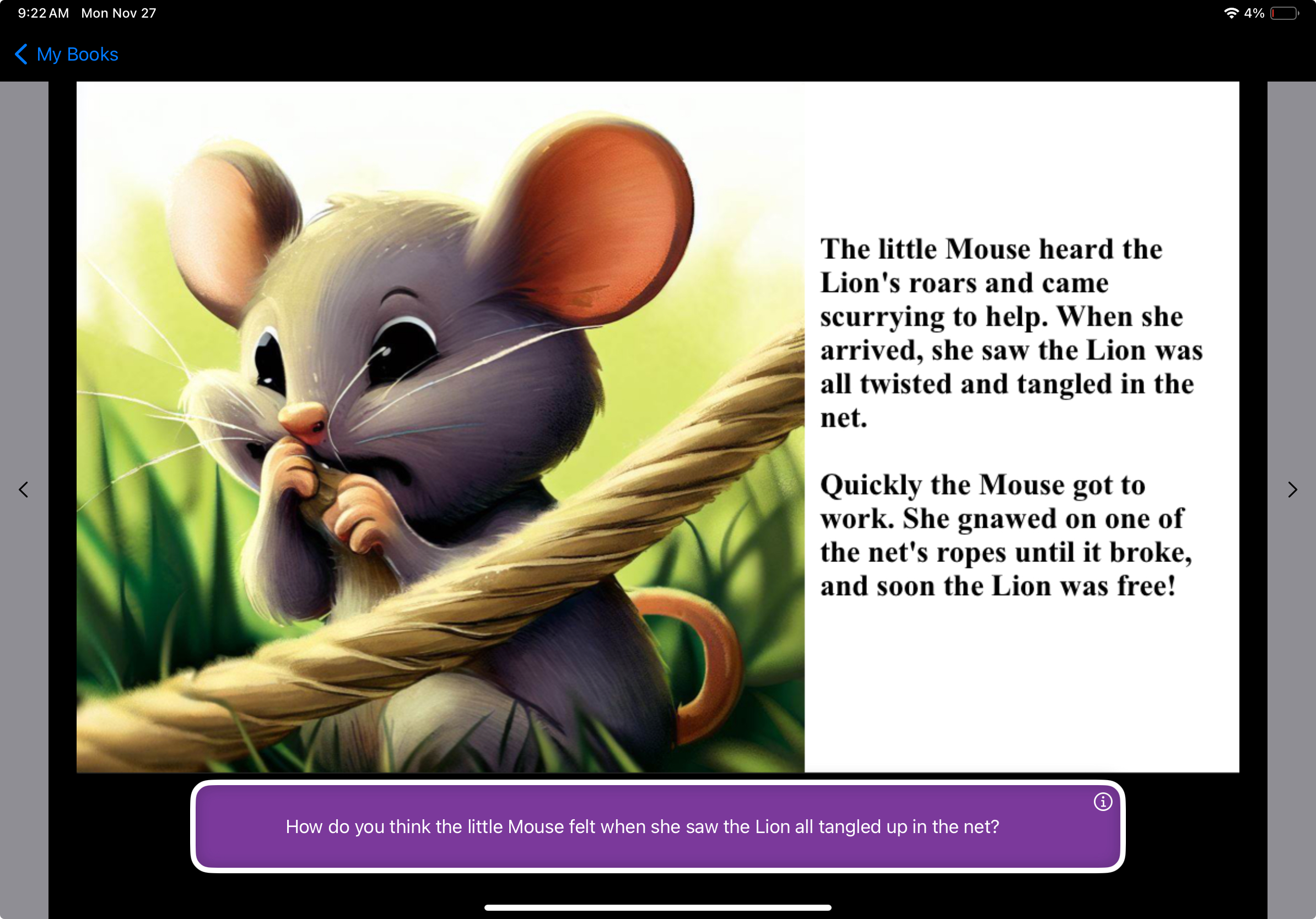}
    \caption{\system{} presents dialogic questions to parents via a tablet-based reading application.}
    \label{fig:reading-interface}
\end{figure*}

\subsection{Co-Reading Interface}
Having created a question generation module that could reliably output suitable dialogic questions, we then turned to developing a user-friendly interface to display these questions in an unobtrusive way to caretakers. Critically, for this prototype, the type of CROWD question generated and presented was randomly selected; because not every question type can be generated on each page (e.g., recall questions cannot appear on the first page and completion prompts are only suitable in instances with rhymes or repetition), if the system tries and fails to generate a question of a certain type three times, it will move onto a different question type.

We designed the tablet-based interface such that the question would appear in a colored tooltip at the bottom of the screen, without covering any of the content on the page (see \Cref{fig:reading-interface}). This tooltip has a small info button in the top right corner that presents a three-sentence pop-up describing the importance of dialogue during co-reading.

\subsubsection{Iterative Design Testing}
To correct bugs and ensure usability, we conducted iterative testing with 8 parent-child groups (seven dyads and one two-child triad) remotely over video conferencing software. The parents were employees of a large technology company with a mix of technical and non-technical roles and their children ranged in age from 4 to 7 years old ($M=6.1$, $SD=1.2$). We selected this age range as one where children would be comprehensible and able to engage in dialogue but not yet reading independently. (In practice, many 7 year olds preferred reading themselves over being read to, hence the narrowed age range in our later evaluation study.)

After giving informed consent, families either installed our prototype application onto their iPad or we screenshared an iPad simulator in cases when installation was not possible (e.g., incompatible software version). We then observed as families read two children's stories---one without on-screen prompts and one with them---before collecting feedback from them on how to improve their experience. For this iterative design testing, we used hard-coded dialogic questions rather than generated ones in order to get feedback on specific questions and question types.

Based on these observations and families' feedback we made several design decisions. First, parents expressed confusion on when the question should be read, so we chose to move the prompt from the top of the screen to the bottom to suggest reading it after finishing the page. Second, because younger children were distracted by the presence of touch points on the screen, we also removed an audio button from the tooltip which played synthesized speech of the prompt out loud. Finally, we implemented a page-turn button in place of page turn swipes to allow families to point at and touch the page during dialogue; we later increased the tap target dimensions of this button based on iterative testing observations of mistaps.

Once we ran several consecutive sessions without further design updates, we proceeded to a system evaluation of the connected interface and question generation model, described in the next section.

%% file: 04-method.tex
\section{Evaluation Study}
With \system{}, we aimed to design a dialogic reading interface to support parents in leading contextual dialogue. We conducted a qualitative study to understand the impact of this system on reading-time conversation and how parents use this system in practice. Through this study we aimed to answer the following research questions:

\begin{enumerate}
    \item How does a dialogic reading interface impact parent-child reading-time conversation as compared to a traditional eBook interface?
    \item How do parents use the dialogic reading interface as a tool in practice?
\end{enumerate}

\subsection{Participants}
As we aimed to study the impact of the system on parents as they read aloud to their child, we recruited 12 parent-child dyads to participate in a one hour online study session. To qualify for the study, families had to speak fluent English and all child participants needed to be 4--6 years old (see Table \ref{tab:demographics} for participant details). To include a range of backgrounds, families were recruited through public schools serving communities with low and mid socioeconomic status (based on median household income; N=5 dyads) and via word of mouth (N=7). Families came from 4 different US states and participated using either family-owned iPads or iPads that we lent to participants for this study. They received a \$50 gift card in exchange for their time.

\begin{table*}[]
\begin{tabular}{p{30pt}p{20pt}p{60pt}p{90pt}p{70pt}p{60pt}}
\hline
\textbf{Family\newline ID} & \textbf{Child Age} & \textbf{Parent\newline Relationship} & \textbf{Typical Co-Reading\newline Minutes per Week} & \textbf{Typical\newline Book Type} & \textbf{Recruitment Avenue} \\ \hline
1 & 6 & Mother & 20--30 & Print & School \\
2 & 5 & Mother & 60--75 & Print & School \\
3 & 5 & Mother & 140 & Print & School \\
4 & 4 & Mother & 30--50 & Print & School \\
5 & 6 & Mother & 60--75 & Print & Word of Mouth \\
6 & 5 & Father & 100 & Print & Word of Mouth \\
7 & 4 & Mother & 120--150 & Print & Word of Mouth \\
8 & 6 & Mother & 0--60 & Print & School \\
9 & 6 & Mother & 140 & Print and eBook & Word of Mouth \\
10 & 6 & Mother & 105--140 & Print & Word of Mouth \\
11 & 5 & Father & 105 & Print and eBook & Word of Mouth \\
12 & 6 & Mother & 110 & Print & Word of Mouth \\  
\hline
\end{tabular}
\caption{Participant demographic information}
\label{tab:demographics}
\end{table*}

\subsection{Procedure}
Sessions were one hour long and were conducted remotely via videoconferencing software. Families were instructed to join the call from a device other than the iPad they would be using for the study.

During the session, we first obtained informed consent before installing the application onto the family's iPad. As the app installed, we asked the family for background information on their typical co-reading habits, including when, how often, and why they choose to read together. 

We then asked families to use the application to read two short stories (each 300 words split across 6 pages), which were included in the app for the study. We selected and modified two of Aesop's fables, \textit{The Lion and the Mouse} and \textit{The Fox and the Stork}, to match in style, complexity level, page count, and word count; the first two families read different stories of similar complexity. While reading the first story, the app displayed a standard eBook interface (No Questions) and during the second story the app also presented a generated question on each page (With Questions). We did not indicate that questions would appear in the second book or provide any instruction about how families should use the presented questions. 

We included the No Questions interface to help us understand each family's typical reading behaviors and to serve as a point of comparison for families when reflecting on the presented prompts. Therefore to minimize potential learning effects that could impact the types of questions parents asked or the conversations families had, we chose to always present this No Questions interface first. Critically, we told families directly at the start of the reading phase that we were \textit{``trying to learn about the conversations families have while reading''} so the changing interface between books (No Questions to With Questions) did not clue the parent into the research objectives halfway through the session. We alternated the order in which the stories themselves were read.

After reading both books, we conducted a semi-structured interview to get feedback from families about their experience using the app, their prior knowledge about dialogic reading, and how they might put dialogic reading into practice in their homes. While we focused on the parent perspective during this interview, children were present and permitted to respond as well. We include a full list of interview questions as well as the text of both stories in our Supplementary Materials.

\subsection{Data Analysis}
While we carefully considered the order in which families used the two versions of the interface and read the two stories, the primary goal of the study was to gauge qualitative reactions to using the dialogic question system rather to experimentally compare system versions. We thus focus our analysis and findings on the qualitative data while presenting quantitative comparisons more briefly.

We collected video and audio recordings of participants from all sessions. These sessions were automatically transcribed via the video conferencing software and those transcriptions were manually corrected by the research team. We applied a semantic and realist thematic analysis to this data, taking a combined deductive and inductive approach \cite{braun2006using}. Our final codebook is included in the Supplementary Materials.

We coded interview data at a per interview question unit, focusing on the goals parents have for reading, the challenges they face when engaging in dialogue, the perceived benefits of the app, and any concerns with the app. 

For the reading phase, we identified segments of contextually meaningful dialogue to analyze how conversation and behaviors change when supported by a dialogic reading interface. In this case, contextual meant the dialogue specifically related to the book reading activity and meaningful meant it further connected to the story's content (i.e., plot, themes, lessons, or pictures). This definition therefore excludes off-topic conversation (e.g., the child asking for a snack) and dialogue about the activity, but not the content (e.g., who presses the page turn button). Given the focus of our research, our unit of analysis for this book-reading phase of the study was a dialogic interaction, which we identified as a conversation during a pause in reading separated from the prior or next dialogic interaction via 1) a page turn, 2) an off-topic conversation, or 3) a clear transition statement (e.g., "Something else I was going to ask..."). 

While we focus our analysis on qualitative findings that convey \textit{how} \system{} influenced reading behaviors, we report a more quantitative perspective of the collected data to explore \textit{how much} of an impact the presented questions had. To quantify dialogue, we consider the number of contextually meaningful conversational turns from both the parent and the child, where a contextually meaningful conversation turn is an utterance by the parent or child related to the book's contents. We then look to individual dialogic interactions to lend additional perspective on the contents of conversations by reporting the prevalence of each CROWD prompt type (i.e., the first question asked in a dialogic interaction) and the number of dialogic interactions covering each identified conversation topic (e.g., plot or morals). Finally we consider how parents used \system{} as a tool in practice by reporting on the number of times parents used, modified, or ignored presented dialogic questions.

%% file: 05-results.tex
\section{Results}
From our thematic analysis, we identified three key themes in this data that describe how family conversations differ when supported by ML-generated dialogic questions, and how such a reading interface might impact family reading practices. 

\subsection{Theme 1: Generated Questions as a Parent Guide for Conversation}
Parent participants described \system{} as a ``guide'' to help them better engage with their child. Parents by and large reported that they do typically try to talk to their child about books while reading them, but they also described the challenges they regularly faced around leading conversation including forgetfulness, exhaustion, or difficulties thinking of questions to ask. By contrast, when reflecting on their experiences using the app, they described how the presented dialogic questions could help them overcome these common challenges.

When discussing their typical home reading behaviors parents identified challenges that often prevent them from starting conversations. Some, like P10, reflected on \textit{``questions that sometimes we forget to ask because we're just reading,''} suggesting that a focus on getting through the story can distract from the goal of dialogue. Other parents reported that even when they remember to ask questions, \textit{``it's kind of hard to think of new ways to make it kind of a conversation,''} especially when \textit{``kids love to reread books over and over and over again''} (P11). 

Not all parents start conversations while reading though, as some described that when reading at home they often consciously choose not to ask questions because they lack the energy or presence to think of them. P11 typically reads to his children right after work, and can find it hard to completely pull his mind away from other tasks: \textit{``If I am worried about something or I've got something brewing in the back of my head it's harder to think or want to make conversation.''} P8, on the other hand, reads to his child at bed time, but says that one downside of this is that \textit{``the questions are less interesting because there's the—there’s just like a tiredness, both of child and adult.''} P3 similarly points to exhaustion as a reason why she may choose to skip conversation while reading, especially when she's had a long day. \textit{``Sometimes it’s—it's hard...when you're at the end of your day, if you've been working all day, to have the questions to say: How are you feeling? What did you notice?''} 

However, parents believed that \system{} would serve as a reminder to talk about stories, an inspiration when thinking of questions, and a tool for tired parents. For the forgetful, P8 appreciates the presence of questions \textit{``that are there to remind me to ask her,''} and P3 explains that the on-screen \textit{``prompts can also have you pause and remember what the point of reading it to them is. It's not just to---to read for them to---to listen. It's for them to actually respond as well.''} Parents like P4 who typically have a hard time thinking of questions to ask appreciated that \textit{``it's not a question that I have to come up with myself to try to spark conversation.''} As P10 reported: \textit{``I like the fact that...they gave you the questions to ask them. That way, you know, you don't really have to think about it. You just, you know, go along with it.''} Finally, when reflecting on exhaustion, P3 also believes that \system{} can provide support for parents \textit{``so that sometimes if you're tired, you don't have to come up with the questions yourself.''} Whether forgetful, distracted, tired, or struggling to think of a prompt, parents appreciated that \system{} was \textit{``kind of a guide [that] could be helpful for parents reading to kids''} (P9). 

With participants describing \system{} as an effective parent guide, we would expect to see more meaningful contextual dialogue when presented with dialogic questions than when reading without them. Indeed, we find that parents and children alike had significantly more meaningful conversational turns when reading with presented generated questions (Parents: Range: $7-74$, $M=30.8$, $SD=23.2$; Children: Range: $3-67$, $M=25.1$, $SD=20.2$) than without (Parents: Range: $0-28$, $M=13.4$, $SD=9.3$; Children: Range: $0-25$, $M=10.6$, $SD=8.8$). A paired-samples Wilcoxon test found this difference was significant for both parents ($z=-3.06$, $p=0.002$, $r=0.88$) and children ($z=-2.94$, $p=0.004$, $r=0.85$).

\begin{figure*}[t]
    \centering
    \includegraphics[width=\textwidth]{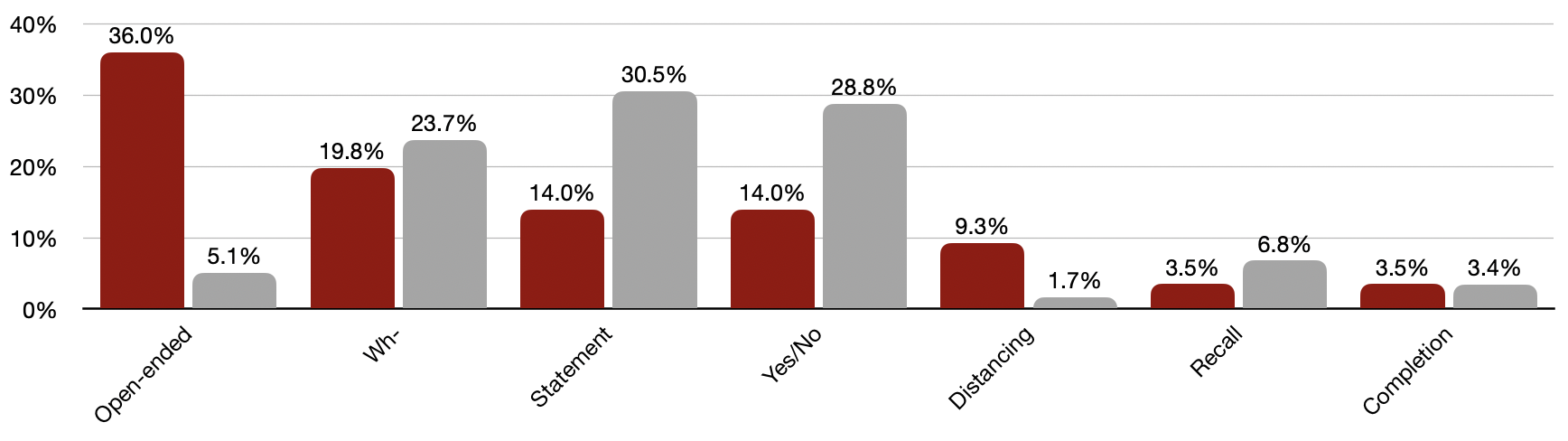}
    \includegraphics[width=\textwidth]{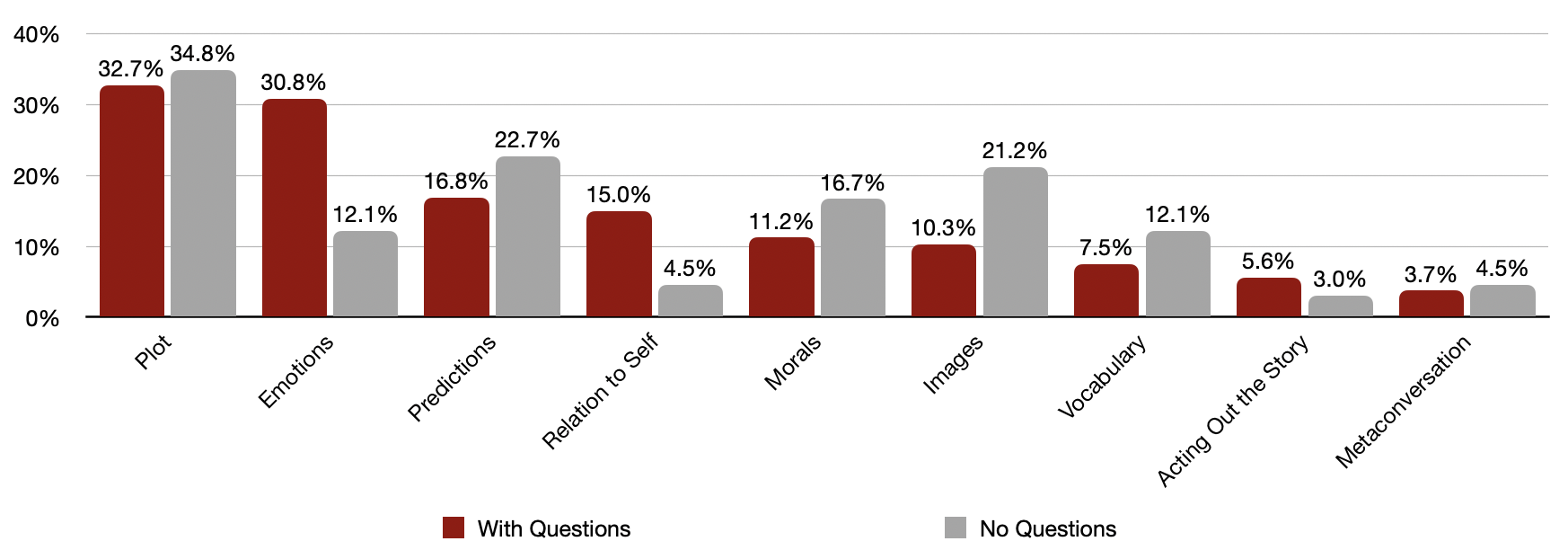}
    \caption{\textbf{Top:} For each interface, the proportion of parent-led dialogic interactions that began with each question type. \textbf{Bottom:} Proportion of dialogic interactions in each interface that touched on a particular topic. Meta conversations were discussions about the reading experience (e.g., do you like the story). Note: conversations could have more than one topic so percentages for each interface do not sum to 100\%.}
    \label{t:prompts-by-condition}
    \label{t:topics-by-condition}
\end{figure*}

\subsection{Theme 2: Depth Beyond Comprehension}
\system{} deliberately presented all five types of CROWD prompts to parents, including the comprehension-based wh- and recall prompt types more commonly found in prior work, as well as the more authentic open-ended and distancing types. We posited that the more authentic questions would lead to richer dialogue, but recognized the need for attention and comprehension checks as well.

Parents reflecting on their reading experience generally reported that they disliked the plot-based comprehension questions so central to prior work because they \textit{``just ask you to just repeat what you just read''} (P6). When discussing which prompts were least useful, P7 specifically called out a wh- prompt that appeared for her while reading: ``What promise did the mouse make to the lion?'' This parent explains that this sort of question doesn't achieve the objectives she has when having dialogue with her child because \textit{``it doesn't really trigger anything except like, you know, rote memorization''} and that asking strictly comprehension-based questions causes her child to tune out: \textit{``Do you know what this means? Do you remember what happens on this page? I frankly think I lose his attention if I do it that way.''} P4 further described how these comprehension questions are less helpful to her when determining how to start a conversation because they're comparatively easy for her to come up with on her own: \textit{``I’m not a very creative person...but, like, asking, like, reading comprehension questions doesn't really require any creativity of me.''}

On the other hand, parents expressed an appreciation for \system{}'s authentic questions that allowed them to achieve a level of depth with their discussions beyond strict comprehension. For example, these questions asked children to consider the character's emotions (e.g., ``How do you think the mouse felt after she helped the lion get out of the net?'') or to relate events in the story to their own life (e.g., ``Have you ever been in a situation where you needed someone's help? How did they assist you?''). When reflecting on how the presented prompts could be most useful, P5 explained: \textit{``I thought the questions easily kind of let you get both characters' perspectives, and also prompted me to go a little bit deeper in what I would have thought to connect the story to...I just feel like these are open, kind of, you know, those open-ended questions that you should be asking.''} More specifically, parents, like P12, wanted to use questions as a way to connect stories to long-running and ongoing conversations with her child: \textit{``It creates a bigger discussion...because I think you can use the books and the stories as, like, life metaphors to just kind of help with, like, life situations. So to really apply it.''} As P2 explains, these questions drive the conversations deeper \textit{``because sometimes I'm not thinking about, like, what the potential lessons are in the book.''}

In particular, parents appreciated being presented with open-ended or distancing questions that tied to lessons they were trying to teach in day-to-day life around morals and emotional awareness. P12 felt the presented questions were \textit{``designed to help pull out like the messages of the story,  create a conversation around the meaning, and help kind of think through the lessons''}, and she felt that this lesson-based questioning \textit{``creates a bigger discussion.''} Notably, it was often the questions related to emotions, in particular, that resonated with parents, because they mirrored ongoing social-emotional lessons the parent was trying to teach at home. P3 has been teaching her child how to express herself and her emotions and thinks guided discussions about books and characters could help with this effort: \textit{``We’ve been actually working a lot on feelings and describing feelings. So I think a lot of the questions were actually helping to discuss feelings.''} Similarly, P7 felt that questions for children related to emotions (e.g., ``What do you think the mouse felt when she saw the lion trapped in the net?'') were \textit{``helpful''} and \textit{``age appropriate''} because \textit{``at this age...especially is when they're trying to understand feelings and things like that.''} 

Parents articulated a clear preference for open-ended and distancing questions over comprehension questions and for discussions about life lessons over discussions about plots. We examined the types of questions parents asked to start conversations and the topics of those conversation (identified through the qualitative coding process) across the No Question and With Question interfaces to study the extent to which \system{} supported this preference. When considering the type of question that parents asked to start a conversation, we find that $45.3\%$ of parent-led conversations started with open-ended or distancing questions when using the With Questions interface; by contrast only $6.8\%$ of conversations with the No Questions interface began with one of those question types (see Figure \ref{t:prompts-by-condition}). Similarly, when looking at topics discussed during these dialogic interactions identified via our inductive coding process, $30.8\%$ of conversations when reading with presented dialogic questions related to emotions as opposed to $12.1\%$ when reading without them. In contrast, other common conversation topics (e.g., plot) were much more balanced between the first and second story (see Figure \ref{t:topics-by-condition}). 

One critical caveat to this finding is that the system itself did not display these question types in equal distribution. For example, each story only had a single page for which a suitable completion prompt could be created, but suitable open-ended prompts are easily created for any page of the story. As a result of this difference and the question type randomization built into the system, a completion prompt appeared on screen only once across the full study, whereas open-ended prompts were displayed 36 times (Recall: 9, Wh-: 12, Distancing: 14). This bias in presentation certainly affected the question type counts above, but simultaneously aligned with parents' articulated preferences.

\subsection{Theme 3: Parents as a Mediator Between the App and the Child}
Finally, we looked at how parents used the presented questions when reading to support conversation with their child and found that parents read the question as written on the screen only about a third of the time (i.e., for 23 of the 72 presented questions). In the remaining instances we observed ways a parent served as a mediator between the app and the child by rewording questions or selectively choosing to disregard them.

A large subset of those rewordings consisted of minor changes where the parent swapped a word or reordered the question, but stayed close to the presented prompt. In 7 cases, a parent split apart a two part question, asking those parts separately, without otherwise changing the wording of the prompt. In an additional 12 cases, the parent slightly reworded the question. Most of these rewordings had minimal affect on the meaning of the prompt (e.g., \textit{``How do you think the lion felt when the mouse ran across his nose?''} became \textit{``How do we think the lion felt after the mouse ran across his nose?''}), but some of the changes added clarity to the prompt itself. For example, P2 prompted their child with the on-screen prompt, \textit{``Can you think of a time when you really wanted to do something but you had to stop yourself? How did you feel?''} However, their child misunderstood the prompt and instead discussed how the character felt during this part of the story. On the following page, the app presented P2 with a similar question (\textit{``Can you think of a time when you really wanted to do something, but you knew shouldn’t? What did you do?''}), but this time P2 chose to reword the question to clarify C2's prior misunderstanding: \textit{``Can you think of a time when you really wanted to do something---you, [C2's name]---but you shouldn’t do it? Like you knew you weren’t supposed to do it.''} 

Aligning with this observed behavior, in the post-session interviews, P3 spoke directly about how she reworded a question for clarity when she felt her child wouldn't understand: \textit{``I think that actually the one about repaying...I don't know that she would actually understand, like, what it means to repay. Um, so maybe to give back.''}

On 7 occasions, we observed instances that we classified as major rewording of the on-screen question. These rewordings appeared to serve one of two purposes. In some cases, the parent drastically simplified the question for the child. For example, P12 took the question \textit{``How do you think the stork felt when they realized they had been tricked by the fox?''} and reworded it to say \textit{``Do you think the stork's going to like that?''} In doing so, she converted an open-ended question into a yes/no prompt. While this change may not lead to as rich of a discussion, a parent may opt to make such a change when the child is getting restless or is unable to articulate an answer to the more complex question.

The other reason we observed question rewording was for instances of malformed questions, which appeared three times throughout the study. We know from our ablation study on the question generation module (see \Cref{s:q-gen-eval}) that not every questions the model outputs is perfectly suitable, so we were interested in how parents would deal with flawed outputs. Twice the system asked how a character was feeling about certain events that would be more appropriately asked about a different character. In both of these cases, the parent mediated between the app and the child to appropriately reword the prompt. For example, P10 corrected a prompt in one such instance while reading it aloud: \textit{``How do you think the stork felt when he saw the fish dinner served in a jar with a---I think it’s how do you think the fox felt?''}

The other case of a malformed question appeared during the second session before we identified a rare edge-case bug that caused a question to show on the incorrect page. In this instance, P2 instructed C2 to disregard the question and turn the page because it could not be answered until later in the story.

Finally, in addition to this one instance of skipping due to a malformed prompt, parents chose to ignore the question on the page 22 other times during the study (i.e., 23 total ignored questions out of 72 presented). On five of those occasions, they chose to ask a different question, whereas on the remaining occasions, they turned the page without asking a question. In some cases, the app prompted parents with two very similar questions in a row; the parent asked the question when it first appeared and then chose not to ask it again on the next page. In other cases, the parents simply reported that they did not initially notice the (deliberately unobtrusive) questions at first, and only began to read them once they had. Finally, parents may choose to ignore questions when they feel the setting and moment is not conducive to meaningful dialogue; as P7 explained, \textit{``depending on the mood we can use it or skip through.''} Ultimately, the on-screen questions are there to support the parent without distracting the child, and it is up to the parent to decide how and when to mediate between their child and the app itself.

%% file: 06-discussion.tex
\section{Discussion}
Through our system and evaluation, we demonstrate that we can generate questions that support meaningful contextual dialogue during parent-child co-reading scenarios. 

In this section, we discuss how generating questions without answers can lead to richer discussion and better support parent preferences, how unobtrusive design supports parents without distracting a child, how the suitability recognizer approach might generalize to other applications, and ways to extend this system in future work.

\subsection{Questions Without Answers for Rich Discussion and Parent Ease}
Through our analysis, we found that parents viewed \system{} as a guide to help them start conversations, leading to significantly more meaningful conversational turns from both parents and children. Parents utilized the suggested questions the majority of the time, but mediated between the app and the child to modify the presented question to fit the context. When studying the contents of those conversations, we learned that parents appreciated open-ended and distancing questions that tied to real-life situations over comprehension questions, because these authentic questions were more difficult to come up with, led to deeper discussion, and provided more opportunities for bonding.

Returning to related work on systems for dialogic questioning, we found a number of prior systems that successfully used on-screen agents to drive dialogue about books \cite{strouse2013effective, troseth2020enhanced, xu2023rosita}, although systems that relied on an adult co-reader for this job previously found that extraneous features distracted from this task and lead to low question utilization \cite{raffle2011hello}. By contrast, with \system{}'s simple interface, we found high question utilization and significant increases in the amount of dialogue.

Prior models and systems that generated questions focused on comprehension questions with predefined answers \cite{zhang2022storybuddy, yao2021ai}. In educational contexts, those questions support students in developing reading comprehension by allowing them to respond and then, critically, check the correctness of the response \cite{zhang2022storybuddy}. When we began this project we established that solely assessing comprehension was not the goal we were striving for. Instead we followed the CROWD method to generate questions that are germane to dialogue in order to encourage parents toward reading practices that have been shown to advance children's early literacy skills. Parents found suggested comprehension questions like those so dominant in prior work comparatively unhelpful. We initially included more authentic open-ended and distancing questions because they support rich discussion, but we found that these were the questions parent preferred the most but felt least equipped to come up with on their own.

Interestingly, this finding reflects more recent research in supporting parent-child conversation outside of reading contexts (i.e., at mealtimes) \cite{leech2018brief, leech2021intervention}. This body of work trains parents in conversations about recalling past events, explaining concepts, or discussing the future, and shows such talk led to more conversation turns \cite{leech2021intervention} and more abstract references in children's everyday language \cite{leech2018brief}. The work we present in this paper contributes to this growing evidence on the importance of and preference for open-ended conversation strategies between parents and children.

\subsection{Generalizing the Suitability Recognizer to Other Applications}
Traditional fine-tuning methods are computationally expensive and require high-quality data that can demand dozens to hundreds of hours of expert time to curate or annotate \cite{bommasani2021opportunities, xu2022fantastic}. We developed the rubric as a way to improve model output quality by encoding expertise in the absence of an appropriate dataset and the suitability recognizer in order to apply that expertise within a system. In this paper, we show that this technique is highly effective while also requiring orders of magnitude less expert time to implement (just 3 hours), no additional computation for training, and no dataset.

One limitation of the specific instantiation of this approach presented in this paper is that our rubric is based on the input of a single expert. While we show that just one expert's input can still yield higher quality generated questions, it is possible that incorporating input from multiple experts in rubric development could yield even better results, albeit at the cost of additional expert time.

While dialogic question generation is one application of this approach, we imagine it could be generalized to other use-cases when expertise is needed to evaluate output quality or when finding large-scale datasets is challenging. For example, considerable work has explored the space of using LLMs for short story generation. Fine-tuning could leverage datasets of stories, but most high-quality stories fall under copyright. Alternatively, developers could consult narratologists to identify what makes a compelling story. Encoding that expertise into a suitability recognizer could provide a filter to identify and correct for high-quality stories. 


\subsection{Future Work}
Looking ahead, we see several opportunities to extend the work presented in this paper.

\subsubsection{Generating Prompts from Images}
\system{} used only the text of a story as input when generating questions. However, during the evaluation study, we observed how families occasionally chose to discuss the images in the story. Furthermore, while we selected stories that were entirely comprehensible without images, some children's books (e.g., \textit{Milo's Hat Trick} by Jon Agee) include pages where the story is conveyed with images alone. Future work might consider leveraging multimodal LLMs to consider images as input for question generation alongside story text.

\subsubsection{Considering Additional Context when Assessing Suitability}
Observations of parents opting to ignore questions due to their similarity to prior questions suggests an opportunity to consider additional context beyond the text on the page. At the simplest, future iterations of this system could include in suitability recognition some check that the generated question is sufficiently semantically different from previously presented questions or allow the family to select which question type should be presented.

Taking this notion of context a step further though, future work may consider the individual child when determining suitability or what type of question to ask. For example, it may make sense for a child reading a book for the fifteenth time to be asked different (e.g., more cognitively complex) questions than a child hearing it for the first time. Similarly, a parent may want to ask different questions to their four year old than to their six year old, even when reading the same story, due to the older child's more advanced vocabulary, narrative understanding, or theory of mind. We witnessed how parents engaged in a mediation process to adjust questions to suit their child, but given the right contextual information, it is possible for a system to take on some of this work as well.

\subsubsection{Longitudinal Evaluation}
Finally, dialogic questioning as a practice has proven so successful across socio-economic groups, cultures, and languages because researchers have demonstrated that caretakers and educators trained in this practice can help children jump months ahead in early literacy skills \cite{whitehurst1988accelerating, valdez1992accelerating, arnold1994accelerating, lim2002facilitating}. We developed \system{}, though, with the understanding that such training does not reach most parents; indeed when asked at the end of the study, only three of the twelve participating parents reported prior formal knowledge of the importance of asking questions to children while reading. Our work shows that machine-generated prompts support untrained parents in leading contextual dialogue while co-reading, but future work should explore if this approach can 1) show quantifiable differences in a large-scale experimental controlled setting, 2) teach parents to lead more consistent or effective dialogue over time, even in the absence of technological support, and 3) lead to childhood literacy gains in a manner that mirrors prior studies on the effectiveness of dialogic reading when introduced with formal training.

%% file: 07-conclusion.tex
\section{Conclusion}
In this work, we presented \system{}, a tablet-based reading application to unobtrusively present dialogic questions to parents during co-reading. We generate these questions by leveraging an LLM in conjunction with a suitability recognizer that encodes educational expertise in a rubric to identify high-quality outputs. This system produces higher-quality questions with orders of magnitude less expert time and no additional training. Through a qualitative evaluation, we find that \system{} serves as a guide for parents leading to more contextually meaningful conversation turns from both the parent and the child, supports parents in guiding deeper conversations that tie to real-life learning objectives through more open-ended and distancing questions, and enables parents to mediate question content and wording between the app and their child. Looking ahead, this work presents exciting opportunities for technology to close gaps in early literacy education by supporting caretakers in making the most of reading time with their children.

%% file: 08-selection-of-children.tex
\section{Selection and Participation of Children}
Families were recruited from an early childhood center in Connecticut, an elementary school in New Jersey, and an elementary school in Texas; additional families were recruited via word of mouth in Texas and California. All contacted families received an email with study information that included a link to contact the research team to express interest in participating. The research team then confirmed eligibility (i.e., within age-range and English speaking), scheduled sessions, and shipped loaner iPads to families as needed for this study. All children provided verbal assent to participate in the study and to be video recorded, and parents signed consent forms prior to video and audio data collection. Families received a \$50 gift card in exchange for participating.

%% file: 09-appendix.tex
\newpage
\onecolumn
\appendix

\section{Rubrics}\label{a:rubrics}

\subsection{Suitability Rubric}
\begin{table}[!h]
\begin{tabular}
{p{.08\textwidth}p{.04\textwidth}p{.1\textwidth}p{.2\textwidth}p{.2\textwidth}p{.28\textwidth}}
\textbf{Type} & \textbf{Level} & \textbf{Authenticity} & \textbf{Grammatical Structure} & \textbf{Relevance to Book} & \textbf{Other Suitability Criteria} \\ \hline
Completion & 1 & Inauthentic & 
\begin{itemize}
\vspace{-3mm}
\item Completion phrase is at most one sentence long
\item Blank should be at the end of the phrase
\end{itemize}
& \begin{itemize}
\vspace{-3mm}
\item Completion phrase is on current page
\end{itemize} & \begin{itemize}
\vspace{-3mm}
\item Deals with rhyming or repeated phrases
\end{itemize} \\ \vspace{0.2cm} \\
Recall & 2 & Inauthentic & 
\begin{itemize}
\vspace{-3mm}
    \item The question is not a composite of multiple questions
    \item Starts with an interrogative adverb/pronoun
\end{itemize}
& \begin{itemize}
\vspace{-3mm}
    \item Asks child to summarize thematically important events
\end{itemize} & \begin{itemize}
\vspace{-3mm}
    \item Asks child to summarize elements of plot or describe sequences of events
    \item Answer cannot be determined solely from the current page
\end{itemize} \\ \vspace{0.2cm} \\
Open-Ended & 3 & Authentic & \begin{itemize}
\vspace{-3mm}
    \item Starts with an interrogative adverb/pronoun
\end{itemize} & \begin{itemize}
\vspace{-3mm}
    \item Questions should relate to story themes, soliciting speculation about or foreshadowing upcoming story events
    \item Relates to the current page of the story
\end{itemize} & \begin{itemize}
\vspace{-3mm}
    \item Solicits ideas or opinions about story elements or asks child to speculate about something related to the story (e.g., plot, characters, setting)
    \item Does not directly ask about child's personal experiences, but child may need to draw on personal experiences to answer
    \item Children should not easily be able to opt-out of answering the question
    \item Question discourages one word answers
\end{itemize} \\ \vspace{0.2cm} \\
Wh- & 1 & Inauthentic & 
\begin{itemize}
\vspace{-3mm}
    \item Start with an interrogative pronoun
    \item Not a composite of multiple questions
\end{itemize} & 
\begin{itemize}
\vspace{-3mm}
    \item Details should pertain to objects or characters that are thematically important to the story plot
    \item Answer is in the text or pictures on the current page
\end{itemize} & \begin{itemize}
\vspace{-3mm}
    \item Focuses on story details
\end{itemize} \\ \vspace{0.2cm} \\
Distancing & 3 & Authentic & 
\begin{itemize}
\vspace{-3mm}
    \item Start with an interrogative adverb/pronoun or a verb
\end{itemize}
 & 
\begin{itemize}
\vspace{-3mm}
    \item Relates to the current page of the story
\end{itemize} & 
\begin{itemize}
\vspace{-3mm}
    \item Explicitly asks the child about their experiences
    \item Relates to the current page
    \item Cannot be answered in one word
\end{itemize}
\end{tabular}
\caption{Question suitability rubric used to determine if a question of a given type generated from the system is suitable for presentation. Level is further described in Appendix \ref{t:levels}.}
\end{table}
\clearpage

\subsection{Question Levels}
\begin{table}[!h]
\begin{tabular}{lp{0.85\textwidth}}
\textbf{Question Level} & \textbf{Description} \\ \hline
Level 1 & Information recall questions focused on what can immediately be seen (or read) in the text. Questions ask students to define, describe, list, or name attributes or utility of objects or characters in the text. \\ \vspace{0.05cm} \\
Level 2 & Open ended questions used to solicit the child's feedback. Questions involve information processing, asking students to analyze, compare, contrast, group, infer, sequence, or synthesize information gathered from the text. \\ \vspace{0.05cm} \\
Level 3 & Questions are related to the story plot, but may also relate to the child's personal experiences or remote events. Questions ask the child to apply, evaluate, hypothesize, imagine, judge, predict, or speculate about the story and their own experiences. \\ \vspace{0.05cm} \\
\label{t:levels}
\end{tabular}
\caption{Levels for rubric suitability as adapted from Costa \& Lowery \cite{costa1989techniques} and Flynn \cite{flynn2011}.}
\end{table}

\section{Prompts}\label{a:prompts}
This appendix section gives example prompts for generating and suitability checking an open-ended question. Words in all caps and square brackets were included verbatim as prompt variables. Words in parentheses were replaced with the relevant piece of text.
\subsection{Generation Prompt}
The initial prompt is used to generate a candidate question.
\begin{lstlisting}[breaklines]
Act as a early childhood reading instructor, producing 'dialogic reading' prompts that encourage conversation and engagement with the text.

Generate an 'openEnded' prompt, that encourages the child to express their own ideas and opinions about the story.

This prompt should allow for creativity and imagination. Avoid questions that can be answered with a simple yes or no.
                    
Remember that the language you use to create [PROMPT] and the themes you pull from the text must be age appropriate for 4-6 year olds.

Make sure [PROMPT] is to the point, and is not verbose.

Read the following text and use it to better understand the characters and events of the main text block. DO NOT use any of the text for prompting.

(previous page text)

"With that context, generate a prompt of type 'openEnded' for this main text: 

(current page text)

Format your response in JSON using exactly the template below:
{
    "prompt": PROMPT
}
        
\end{lstlisting}

\subsection{Suitability Check Prompt}
Once a candidate prompt is generated, suitability is determined through a series of prompts that capture items in the suitability rubric (see Appendix \ref{a:rubrics}). If a prompt is deemed unsuitable, a feedback line is included within the generation prompt text for the next query that includes the unsuitable prompt and the explanation. Here is an example suitability prompt that determines authenticity.

\begin{lstlisting}[breaklines]
Act as a early childhood reading instructor. You will be judging if [PROMPT] is [AUTHENTIC], given the [CURRENT_PAGE] and [PREVIOUS_PAGES] of the story book as context.
Remember that [AUTHENTIC] is defined as follows:
True if [PROMPT] does NOT have a prescribed answer on [CURRENT PAGE] or [PREVIOUS PAGES].
False if [PROMPT] has a prescribed answer, which can be determined from [CURRENT PAGE] or [PREVIOUS PAGES].

With all of this in mind, please help define the [AUTHENTICITY] of [PROMPT]:
    Format responses in JSON using exactly the template below:
        {
            "Authentic" : {[AUTHENTIC]'s value, one of True or False},
            "Explanation" : {Explanation of why [AUTHENTIC] was chosen},
        }

[PREVIOUS PAGES] : (text of previous 5 pages)
[CURRENT PAGE] : (current page text)
[PROMPT]: (prompt)
\end{lstlisting}